\documentclass[english,twocolumn,aps,prl,showpacs]{revtex4-1}

\usepackage{graphicx} 
\usepackage{amssymb}
\usepackage{amsmath}
\usepackage{phonetic}
\usepackage{babel} 



\begin{document}

\title{Thermodynamics of an attractive 2D Fermi gas}

\author{K. Fenech, P. Dyke, T. Peppler, M. G. Lingham, S. Hoinka, H. Hu, and C. J. Vale$^{*}$} 
\affiliation{Centre for Quantum and Optical Sciences, Swinburne University of Technology, Melbourne 3122, Australia 
\\{$^\ast$To whom correspondence should be addressed; E-mail: cvale@swin.edu.au}}

\date{\today}

\begin{abstract}

Thermodynamic properties of matter are conveniently expressed as functional relations between variables known as equations of state. Here we experimentally determine the compressibility, density and pressure equations of state for an attractive 2D Fermi gas in the normal phase as a function of temperature and interaction strength. In 2D, interacting gases exhibit qualitatively different features to those found in 3D. This is evident in the normalized density equation of state, which peaks at intermediate densities corresponding to the crossover from classical to quantum behaviour.

\end{abstract}


\maketitle

Two-dimensional (2D) quantum matter can display behaviors not encountered in three-dimensions (3D) \cite{mermin66,hohenberg67}. Fermions confined to 2D planes play a key role in unconventional superconductors \cite{dagotto94}, graphene \cite{geim07} and certain topological insulators \cite{hasan10} yet understanding the properties of these complex materials can present significant theoretical challenges. Simpler systems, such as ultracold atomic gases with tunable interactions \cite{bloch08,bcsbecbook12,levinsen15}, can serve as valuable testbeds for building up and validating models of interacting fermions in 2D. Experiments on 2D Fermi gases to date have addressed their production \cite{modugno03,martiyanov10,dyke11}, pairing \cite{frohlich11,sommer12}, pseudogap \cite{feld11} and polaron physics \cite{koschorreck12,zhang12}, pair condensation \cite{ries14}, the Berezinskii-Kosterlitz-Thouless transition \cite{murthy15} and the pressure \cite{makhalov14} as a function of interaction strength at low temperatures. Theoretically, several groups have investigated superfluidity \cite{petrov03,martikainen05,botelho06,zhang08,fischer14} and the thermodynamic properties of 2D Fermi gases \cite{bertaina11,bauer14,anderson15}, however, a full experimental characterization remains to be established. 

In cold atom experiments, a 2D gas can be produced by subjecting a cloud to tight transverse confinement. In a harmonic potential, with characteristic frequencies $\omega_x, \omega_y$ and $\omega_z$, a gas can become kinematically 2D when the transverse ($z$) confinement energy exceeds all other energy scales including the thermal energy, chemical potential and interaction energies. The first two criteria require that $\hbar \omega_z \gg k_B T, E_F$ where $\hbar$ is Planck's constant, $k_B$ is Boltzmann's constant, $T$ is the temperature and $E_F$ is the Fermi energy. The criterion for the interactions requires that neither elastic or inelastic (e.g. pair-forming) collisions result in the population of transverse excited states \cite{petrov01,idziaszek05,kestner06,haller10,sala12,levinsen15}. At nonzero densities it was found that moderate interactions can lead to transverse excitations, even when $k_B T$ and $E_F$ lie below the transverse oscillator energy \cite{dyke14}. This in turn can impact other parameters such as the pairing gap \cite{fischer13} and superfluid transition temperature \cite{fischer14}.

Here we measure the thermodynamic properties of attractive Fermi gases in the normal phase in a parameter regime where 2D kinematics has been clearly established \cite{dyke14}. We apply a protocol based on the approach used for a 3D Fermi gas at unitarity \cite{ku12} and employed on a 2D Bose gas \cite{desbuquois14}. By establishing a model independent relationship between the compressibility and pressure, we extract a range of thermodynamic properties including the temperature, chemical potential and density equations of state. 

In a two-component Fermi gas thermodynamic variables can be expressed as functions, $f$, connecting the underlying energies within the system \cite{ho04}. At fixed temperature and interactions, these can be related to the density via the Gibbs-Duhem equation. In 2D the pressure $P$, density $n$ and isothermal compressibility $\kappa$ are given by 
\begin{equation} \label{eq:1} 
P = \frac{1}{\beta \lambda^2}  f_P(\beta \mu, \beta E_b) = \int_{-\infty}^\mu n(\mu',T) d\mu' \; \Big{|}_{T, a_{2D}}
\end{equation} 
\begin{equation} \label{eq:2} 
n \lambda^2 = f_n(\beta \mu, \beta E_b) \end{equation} \begin{equation} \label{eq:3} \kappa = \frac{\beta}{(n \lambda)^2}  f_{\kappa}(\beta \mu, \beta E_b) = \frac{1}{n^2} \frac{d n(\mu,T)}{d \mu} \; \Big{|}_{T, a_{2D}} 
\end{equation}
where $f_i(\beta \mu, \beta E_b)$ depend on the chemical potential $\mu$, temperature and interaction strength, $\beta = 1/(k_B T)$, $\lambda = \sqrt{2 \pi \hbar^2 / (m k_B T)}$ is the thermal de Broglie wavelength, $m$ is the atomic mass and $E_b$ is the two-body binding energy which in quasi-2D is governed by $\omega_z$ and the 3D scattering length $a$. The 2D scattering length $a_{2D}$ is related to the binding energy by $E_b = \hbar^2 / (m a_{2D}^2)$ in the kinematically 2D regime \cite{petrov01,bloch08,levinsen15}. Knowledge of the functions $f_i$ represents a complete determination of the thermodynamics and can establish valuable benchmarks for comparison with many-body theories \cite{vanhouke12}. 

\begin{figure}[!t] \centering
\includegraphics[clip,width=0.48\textwidth]{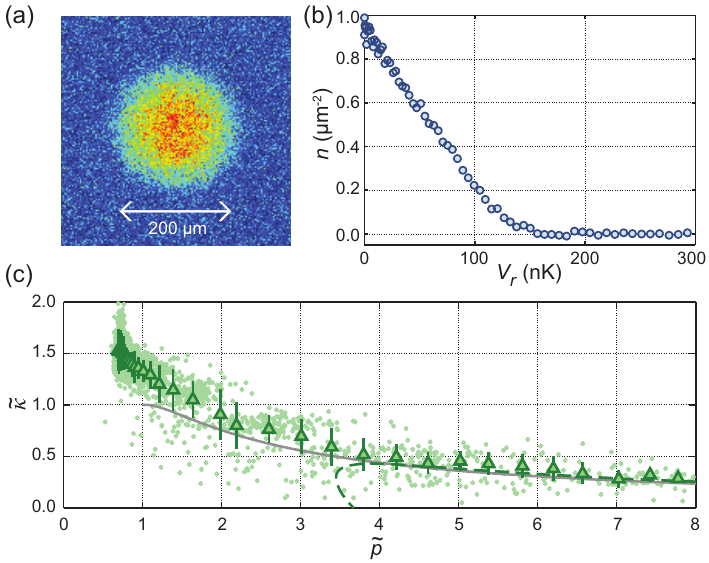} \caption{(a) Average of 10 absorption images of a 2D Fermi gas, with $\beta E_b$ = 0.26, prepared under the same experimental conditions ($B$ = 880 G, $N$ = 16,000, $T \approx$ 25 nK) (b) Azimuthally averaged density, $n(V_r)$, obtained from image (a) as a function of the local potential. (c) Dimensionless compressibility $\tilde{\kappa}$ vs.~dimensionless pressure $\tilde{p}$ of a 2D Fermi gas with $\beta E_b$ = 0.26(2). Faint green circles show data extracted from single images and dark green triangles are data binned according to $\tilde{p}$. Grey solid line shows the equation of state for a noninteracting Fermi gas and dashed green line is the predicted $\tilde{\kappa}$ based on the virial expansion to third order.} 
\label{fig:fig1}
\end{figure}

In the experiments that follow we study 2D atom clouds at thermal equilibrium in a cylindrically symmetric harmonic trap $V_r(x,y) = m \omega_r^2 (x^2+y^2)/2$ with $\omega_r (= \omega_x = \omega_y) \ll \omega_z$. Due to the slowly varying radial potential we can make use of the local density approximation (LDA) which asserts that local thermodynamic properties will be equivalent to those of a homogeneous gas at the same temperature and chemical potential \cite{dalfovo99}. Under the LDA the atomic density can be written as $n[\mu(x,y), T]$, where $\mu(x,y) = \mu_0 - V_r(x,y)$ and $\mu_0$ is the chemical potential at the trap center. In any single realisation of the experiment, the parameters $\beta$ and $E_b$ will be fixed across the cloud such that Eqs.~(1-3) yield $f_n(\beta \mu, \beta E_b) = f_P'(\beta \mu, \beta E_b)$ and $f_{\kappa} (\beta \mu, \beta E_b) = f_P''(\beta \mu, \beta E_b)$ by differentiation with respect to $\beta \mu$. 

We prepare single 2D clouds of neutral $^6$Li atoms in a hybrid optical/magnetic trap at temperatures in the range of 20-60$\,$nK \cite{dyke14,smith05,rath10,supplement}. A blue-detuned TEM$_{01}$ mode laser beam provides tight confinement along $z$ with $\omega_z/(2 \pi) = 5.15 \,$kHz. Radial confinement arises from residual magnetic field curvature present when the Feshbach magnetic field is applied, leading to a highly harmonic and radially symmetric potential with $\omega_r / (2 \pi) = 26 \,$Hz with an anisotropy of less than $0.6 \, \%$. We image along $z$ to directly obtain the density $n(x,y)$ of clouds prepared in the kinematically 2D regime. Fig.~1(a) shows the average of 10 images taken under identical experimental conditions. Due to the symmetry of $V_r(x,y)$, and a precise calibration of the absorption imaging \cite{supplement}, we can azimuthally average these images to obtain $n(V_r)$ as shown in Fig.~1(b). 

\begin{figure}[!t] \centering
\includegraphics[clip,width=0.48\textwidth]{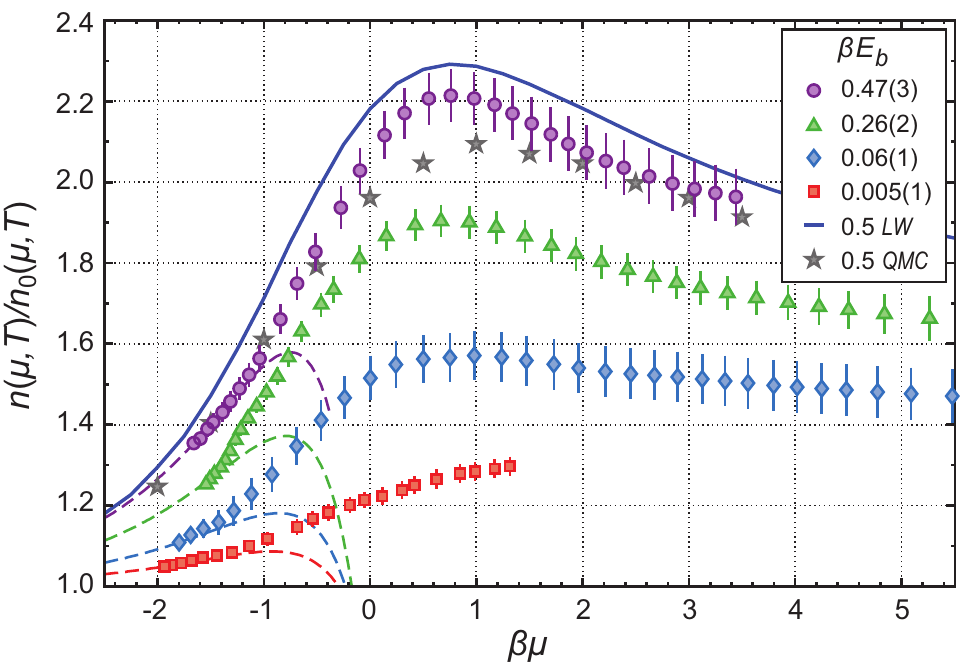}
\caption{Normalized density equation of state for a 2D attractive Fermi gas for $\beta E_b$ = 0.47(3) (purple circles), $\beta E_b$ = 0.26(2) (green triangles), $\beta E_b$ = 0.06(1) (blue daimonds) and $\beta E_b$ = 0.005(1) (red squares). Dashed lines show the calculated equation of state using the third order virial expansion \cite{liu10}. Solid blue line and grey stars are the calculated equation of state for $\beta E_b$ = 0.5 based on Luttinger-Ward (LW) \cite{bauer14} and quantum Monte-Carlo calculations (QMC) \cite{anderson15}, respectively. }
\label{fig:fig2} 
\end{figure}

Both $a$ and $V_r(x,y)$ are precisely known for a given experimental sequence providing accurate knowledge of $E_b$ and, via the LDA, the change in chemical potential $d \mu (\equiv - d V_r)$ across the cloud. Other parameters however, including the absolute temperature and chemical potential, are unknown. Furthermore, absorption imaging is susceptible to systematics that make precise calibration of the atom density challenging  \cite{reinaudi07,yefsah11,chomaz12,hung13}. To proceed we first obtain an estimate of the absolute temperature and chemical potential by fitting the low density wings of the cloud with the virial expansion in 2D to third order \cite{liu10}. As only a small fraction of the cloud can be used, this can lead to relatively large uncertainties in the fits \cite{supplement}. Fortunately, the relationships connecting $f_p, f_n$ and $f_{\kappa}$ allow this to be improved. In analogy with refs \cite{ku12,desbuquois14}, we use the $n(V_r)$ data to construct a model independent equation of state for the dimensionless compressibility $\tilde{\kappa} = \kappa / \kappa_0$ versus pressure $\tilde{p} = P/P_0$ \cite{ku12} where $\kappa_0 = 1/(n E_F)$ and $P_0 = n E_F/2$ are the local compressibility and pressure of an ideal Fermi gas at $T = 0$, respectively, $E_F = (\hbar^2 \pi/m) n = k_B T_F$ is the local Fermi energy and $T_F$ is the Fermi temperature. In Fig.~1(c) we plot $\tilde{\kappa}$ against $\tilde{p}$ for $\beta E_b$ = 0.26(2). At high temperatures the compressibility lies close to that for the ideal gas yet it deviates above this at lower temperatures (lower $\tilde{p}$). The virial expansion provides a reliable prediction of $\tilde{\kappa}$ for $\tilde{p} \gtrsim 6$. Eqs.~(1)-(3), along with $d \mu = -dV_r$, then allow one to find the relative temperature $\tilde{T} = T/T_F$ and $\beta \mu$ at any position in the cloud using the integrals 
\begin{equation} \label{eq:4} 
\tilde{T} = \tilde{T}_i \exp{ \left [ \frac{1}{2} \int_{\tilde{p}_i}^{\tilde{p}} \frac{d \tilde{p}'}{\tilde{p}'-\frac{1}{\tilde{\kappa}}} \right ]} 
\end{equation}
\begin{equation} \label{eq:5} 
\beta \mu = (\beta \mu)_i - \int_{\tilde{T}_i}^{\tilde{T}} \frac{1}{\tilde{T'}^2} \left(\frac{1}{\tilde{\kappa}} \right ) d \tilde{T'} 
\end{equation} 
where the initial points can be chosen to lie in the range where the virial expansion is accurate. Implementation of Eq.~(4) turns out to be relatively insensitive to the precise starting conditions and provides highly robust thermometry directly from the $\tilde{\kappa}$ vs.~$\tilde{p}$ equation of state. As a validation of the imaging calibration, the absolute temperature found from Eq.~(4) should be consistent across the entire cloud and should match the value of $\beta E_b$ that gave the best fit in the cloud wings using the virial expansion \cite{supplement}.  

The integral for the chemical potential, Eq.~(5), should also yield values of $\beta \mu$ that are consistent with temperature found using Eq.~(4). Additionally, $\mu$ should vary according to the LDA as a function of position across the cloud. Meeting all of these conditions requires accurate calibration of the absorption imaging as full consistency is only obtained with the correct parameters \cite{supplement}. Agreement with the virial expansion at high temperatures provides a complementary validation of the thermodynamic parameters that assures the data is free from systematics within the error bounds of the virial fit.

With $T/T_F$ and $\beta\mu$ known across the cloud, and the above criteria satisfied, we can construct the homogeneous density equation of state. In Fig.~2 we plot $n(\mu, T)$ for four values of $\beta E_b$, normalized to the density of an ideal Fermi gas at the same chemical potential and temperature $n_0(\mu, T) = (2/\lambda^{2}) \ln(1 + e^{\beta \mu})$. Reducing dimensionality leads to a dramatic difference in this equation of state compared to the 3D attractive Fermi gas \cite{nascimbene10,ku12}. The normalized density $n/n_0$ displays a non-monotonic behaviour as a function of $\beta \mu$. This is most evident for the gas with strongest interactions $\beta E_b = 0.47$ where $n/n_0$ peaks just below $\beta\mu = 1$. Also shown are the calculated equations of state using the virial expansion (dashed lines) \cite{liu10}, which is valid for $\beta \mu \lesssim -1.5$, along with recent Luttinger-Ward (LW, solid line) \cite{bauer14} and quantum Monte-Carlo (QMC, grey stars) \cite{anderson15} calculations for a cloud with $\beta E_b = 0.5$. Our data for $\beta E_b = 0.47$ shows good qualitative agreement with these calculations lying in between the LW and QMC curves. 

The peak in $n/n_0$ originates from the presence of a two-body bound state which exists for arbitrarily weak attraction in 2D. Interactions will generally be most significant when the kinetic energy of colliding atoms, $E_k$, is approximately equal to $E_b$. In 2D, when $T \ll T_F$, this occurs when the interaction parameter $\ln{(k_F a_{2D})} \rightarrow 0$ \cite{levinsen15}, where $k_F = \sqrt{2 \pi n}$ is the Fermi wavevector. At $T = 0$, the existence of a bound state leads to the possibility of tuning from the fermionic (BCS) regime where $E_b \ll E_F$, to the bosonic regime with $E_b \gg E_F$, simply by varying the density. At finite temperatures, the thermal energy $k_B T$ sets a lower bound on the collision energy and the Bose limit is only possible when $\beta E_b > 1$. For the data in Fig.~2, $\beta E_b$ is always less than unity and $E_k$ always exceeds $E_b$. In the low density (classical) region of the clouds, where $\beta \mu < 0$, $E_k$ will be set by $k_B T$, and interactions become stronger as the density begins increasing and the relative temperature $\tilde{T}$ decreases. However, once $\beta \mu \gtrsim 1$, $E_F$ becomes the dominant energy (quantum regime) and $E_k$ increases above $k_B T$, and hence further above $E_b$. This leads to effectively weaker interactions at high density and $n/n_0$ begins decreasing for $\beta \mu \gtrsim 1$.

\begin{figure}[!h] \centering
\includegraphics[clip,width=0.482 \textwidth]{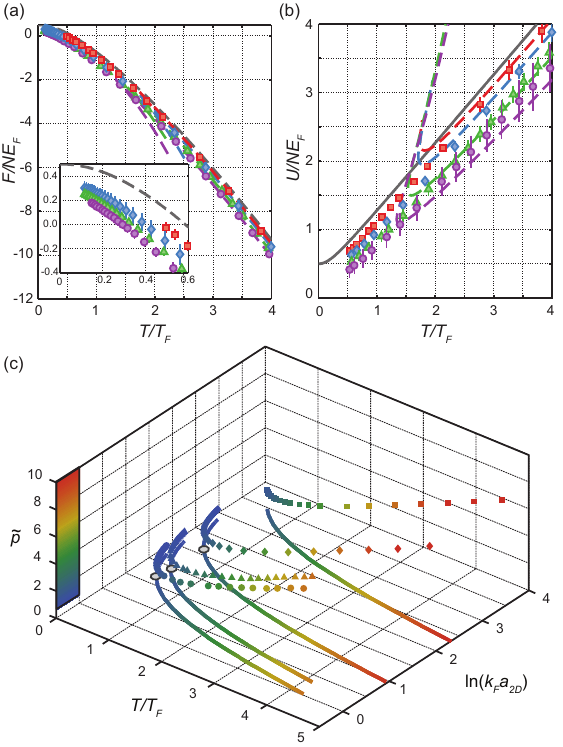}
\caption{(a) Normalized free energy $F$ and (b) internal energy $U$ as a function of $T/T_F$ for $\beta E_b$ = 0.47(3) (purple circles), $\beta E_b$ = 0.26(2) (green triangles), $\beta E_b$ = 0.06(1) (blue daimonds) and $\beta E_b$ = 0.005(1) (red squares). Inset in (a) shows a zoomed in view of $F$ at low temperature. Dashed lines show the calculated energies using the virial expansion for the same values of $\beta E_b$ and grey lines show the ideal gas result. (c) Normalized pressure $\tilde{p}$ as a function of the interaction parameter $\ln{(k_F a_{2D})}$ and temperature $T/T_F$. Solid lines show how the interaction parameter varies as a function of the relative temperature within single clouds and grey circles on the contours indicate the approximate location of the peak in $n(\mu,T)/n_0(\mu,T)$.}
\label{fig:fig3} 
\end{figure}

Not all thermodynamic properties of an interacting 2D Fermi gas can be obtained from measurements on a single cloud. While the free energy, $F = U - TS$, where $U$ is the internal energy and $S$ is the entropy, is readily found via the grand potential $\Omega = - P A = F - \mu N$, where $A$ is the area and $N$ is the particle number, $U$ and $S$ cannot be found individually without further information. The normalized free energy is given by $F/(N E_F) \equiv \tilde{F} =\mu / E_F - \tilde{p}/2$, and is plotted in Fig.~3(a). However, unique determination of $U$ and $S$ requires differentiation across clouds with different $a_{2D}$. Specifically, by considering $\tilde{F}$ as a function of temperature and $a_{2D}$ we can evaluate the contact density ${\cal C}$ using $d \tilde{F}/d(\ln{a_{2D}}) = 2 {\cal C} / k_F^4$ at fixed $T/T_F$. With this the internal energy, and hence the entropy, can be found via the Tan pressure relation \cite{hofmann12,werner12,valiente12}, $\tilde{p} = 2 U/(N E_F) + 2 {\cal C} / k_F^4$ \cite{supplement}. As we only have measurements at four values of $a_{2D}$ the measured contact contains significant uncertainty compared to other variables. However, as ${\cal C}$ remains relatively small in our experiments, the internal energy shows the expected behaviour, Fig.~3(b).

Finally, we can plot the dimensionless pressure $\tilde{p}$ versus interactions and temperature in Fig.~3(c). Constructing a full 3D surface plot of $\tilde{p}$ as a function of $\ln{(k_F a_{2D})}$ and $T/T_F$ would represent a complete characterisation of the thermodynamics of the 2D Fermi gas. The contours in the $\tilde{p} = 0$ plane show how the relative temperature and interaction strength evolve in a single cloud. Grey circles on the contours indicate the approximate peak location in the density equation of state.

In summary we have measured the thermodynamic properties of a 2D Fermi gas with attractive interactions in the normal phase. The existence of a bound state leads to qualitatively different behaviour to 3D gases as apparent in the density equation of state. For the gas with strongest interactions ($\beta E_b = 0.47$) the superfluid transition temperature is expected to lie at approximately $0.05 \, T/T_F$ \cite{bauer14} which is around a factor of two colder than we currently achieve. Future studies investigating thermodynamic signatures of the superfluid transition may provide insight into the nature of the transition in a quasi-2D Fermi gas. At stronger interactions one could investigate the effects of transverse excitations on the thermodynamic properties.

We thank P. Hannaford, J. Drut, M. Parish and J. Levinsen for helpful discussions and comments on the manuscript and the authors of refs \cite{bauer14,anderson15} for sharing their data.  C. J. V. and P. D. acknowledge financial support from the Australian Research Council programs FT120100034, DP130101807 and DE140100647.

{\it Note added}: A related work has recently been posted which examines the thermodynamic equation of state across the 2D BEC-BCS crossover \cite{boettcher15}.

\clearpage
\newpage

\pagenumbering{arabic}

\onecolumngrid

\begin{centering}

\textbf{\large Supplemental Material: Universal thermodynamics of an attractive 2D Fermi gas} \\
\bigskip

K. Fenech, P. Dyke, T. Peppler, M. G. Lingham, S. Hoinka, H. Hu, and C. J. Vale$^{*}$ \\
\smallskip
\textit{\small Centre for Quantum and Optical Sciences, Swinburne University of Technology, Melbourne 3122, Australia \\
$^\ast$To whom correspondence should be addressed; E-mail: cvale@swin.edu.au}

\bigskip
\bigskip

\end{centering}

\twocolumngrid

\setcounter{figure}{0}
\renewcommand{\thefigure}{S\arabic{figure}}

\subsection{Preparing 2D Fermi gases}

To produce a 2D Fermi gas we begin with a 3D gas containing $N/2 = N_\sigma = 2 \times 10^5$ Li$^6$ atoms confined in a red-detuned (1075 nm) single beam optical dipole trap. The cloud is evaporatively cooled to a temperature of $T \sim 0.1 \, T_F$ at unitarity (832.2 G). We then ramp the magnetic field to the BEC side of the Feshbach resonance (780 G) and transfer the atoms to a TEM$_{01}$ mode optical trap produced by a 532 nm frequency-doubled fibre laser \cite{sdyke14}. This provides tight confinement in the transverse direction, while radial confinement is provided by residual curvature in the magnetic field generated by the Feshbach coils. Initially, the TEM$_{01}$ mode laser is ramped up to approximately 10\% of its full power over 350 ms and the 3D optical trap is simultaneously switched off. We then perform evaporative cooling in the weak quasi-2D trap by applying a transverse magnetic field gradient over a period of 2 seconds. Including this extra step makes it possible to achieve lower temperatures and provides precise control over the atom number in the final 2D trap. Lastly, the magnetic field gradient is removed and the TEM$_{01}$ mode is ramped to full power to achieve a tight trapping frequency of $5.15 \pm 0.24$ kHz and a radial trapping frequency of $26.4 \pm 0.1$ Hz at a magnetic field of 972 G. After an equilibration time of 250 ms an {\it in-situ} absorption image is acquired.

\subsection{Calibration of the absorption imaging}

Absorption imaging of the trapped clouds is achieved by illuminating the atoms with a short pulse of resonant laser light propagating along the $z$-direction. The atoms leave a shadow in the laser light which is imaged onto a CCD camera. The atomic density in the $x$-$y$ plane can be inferred from the resulting absorption image via the Beer-Lambert formula including saturation \cite{sreinaudi07} 
\begin{align}\label{eq:density}
    n(x,y) = \frac{1}{\sigma} \left ( \ln \left [ \frac{I_f(x,y)}{I_0(x,y)} \right ] + \frac{I_0(x,y)-I_f(x,y)}{I_{sat}} \right )
\tag{S1}
\end{align}
where $I_f(x,y)$ is the measured intensity with the atoms present, $I_0(x,y)$ is the measured intensity in the absence of atoms, $\sigma$ is the absorption cross-section and $I_{sat}$ is the saturation intensity of the transition. Images used for the thermodynamic analysis are taken with a pulse length of 10 $\mu$s and intensity $I \sim 0.25 \, I_{sat}$. However, with these imaging parameters, the measured density generally underestimates the true density due to the Doppler shift of atoms accelerated by the imaging light and possible multiple scattering at finite gas density \cite{schomaz12}.

\begin{figure}[h!]
\includegraphics[clip,width=0.45\textwidth]{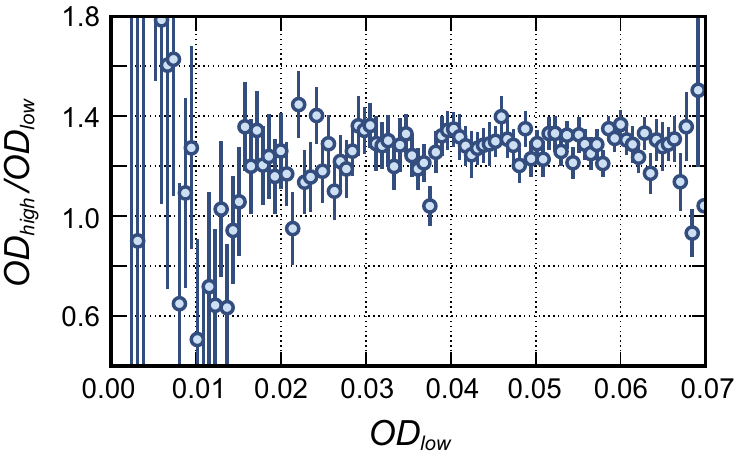}
\caption{Measured optical density correction factor $C_{OD}$ determined at a magnetic field of 972 G. Error bars represent the standard error determined for each bin of $OD_{low}$.}
\label{fig:supp_od_correction}
\end{figure}

To correct for these effects we acquire a series of additional images of the 2D clouds using a 1 $\mu$s imaging pulse with an intensity $I \sim 10I_{sat}$. In this high intensity regime the second term in equation (\ref{eq:density}) dominates and any Doppler shifts during this short pulse are negligible compared to the power broadened width of the transition. The measured optical density ($OD$) under these conditions provides a robust measure of the absolute density \cite{sreinaudi07,syefsah11}, but contains significantly more statistical noise. By averaging many images taken in both low and high intensity regimes of clouds prepared in an identical way we can determine an overall correction to the optical density by constructing a correction factor $C_{OD} = OD_{high}/OD_{low}$. In general, this correction will be a function of the measured optical density, however, as we only ever deal with very low optical densities (typically less than 0.1), $C_{OD}$ turns out to be effectively constant over the range we access. In Fig.~\ref{fig:supp_od_correction} we plot the ratio $OD_{high}/OD_{low}$ versus the measured $OD_{low}$ in the region of the image where atoms are present. As can be seen the correction factor is approximately constant in the range of optical densities measured and becomes noisy at low OD (below $\sim 0.02$).

Averaging the measured correction factor in the range $0.02 < OD_{low} < 0.06$ and applying this to the images provides a corrected density distribution where the uncertainty in the corrected density is typically of order $\sim 10\%$. The actual correction factors determined in this way for the four magnetic fields investigated in the paper are shown in the central column of Table \ref{tab:supp_od_correction}. In the next section, we show how fitting the density using the virial expansion provides a way to improve the accuracy of these coefficients.

\subsection{Temperature estimation using the 2D virial expansion}

\label{sec:temperature_estimate}

The thermodynamic properties of an interacting 2D Fermi gas at relatively high temperatures $T/T_F \gtrsim 1$, where $T_F$ is the Fermi temperature, are well approximated using the virial expansion \cite{sliu10}. When the fugacity $\textit{\yogh} = e^{\beta \mu}$ is a small parameter, one can expand the grand potential in powers of the fugacity, where the expansion coefficient for the $n$-th order term depends on the solution of the $n$-body problem for the interacting particles. For a 2D gas in a harmonic trap the density will be given to third order by 
\begin{align}
    n(\mathbf{r}) = \frac{2}{\lambda^{2}} {\big ( } \ln\left[1+\textit{\yogh}(\mathbf{r})\right]+2\Delta b_{2}\textit{\yogh}(\mathbf{r})^{2}+3\Delta b_{3}\textit{\yogh}(\mathbf{r})^{3}+\cdots {\big ) }
\tag{S2}
\label{eq:virial_fit}
\end{align}
where $\lambda = \sqrt{2 \pi \hbar / (m k_B T)}$ is the thermal de Broglie wavelength, $m$ is the mass of the atom, $\textit{\yogh}(\mathbf{r})  = e^{\beta [ \mu_0-V(\mathbf{r}) ] }$ is the local fugacity, $\mu_0$ is the chemical potential at the centre of the cloud, $V(\mathbf{r}) = m \omega_r^2(x^2 + y^2)/2$ is the trapping potential, $\beta = 1 / (k_B T)$, $\omega_r$ is the radial harmonic confinement frequency and $\Delta b_2$ and $\Delta b_3$ are the 2nd and 3rd virial coefficients, respectively. The coefficients $\Delta b_i$ are functions of $\beta E_b$ where $E_b$ is the two-body binding energy that depends upon the transverse confinement frequency $\omega_z$ and the 3D scattering length $a$ \cite{spetrov01,sbloch08}. $E_b$ is generally known precisely as both the trapping frequency and 3D scattering length are readily determined. 

The second order virial coefficient $\Delta b_{2}$ can be conveniently calculated according to the Beth-Uhlenbeck formalism. By introducing a dimensionless parameter 
\begin{equation}
x=\ln\left(\frac{\lambda}{a_{2D}}\right)=\frac{1}{2}\ln\left(2\pi\beta E_{b}\right),
\tag{S3}
\end{equation}
we obtain 
\begin{equation}
\Delta b_{2}=e^{\beta E_{b}}-\int\limits _{-\infty}^{\infty}dt\exp\left[-\frac{e^{2t}}{2\pi}\right]\frac{2}{\pi^{2}+4\left(t-x\right)^{2}}.
\tag{S4}
\end{equation}

\begin{figure}
\begin{centering}
\includegraphics[clip,width=0.45\textwidth]{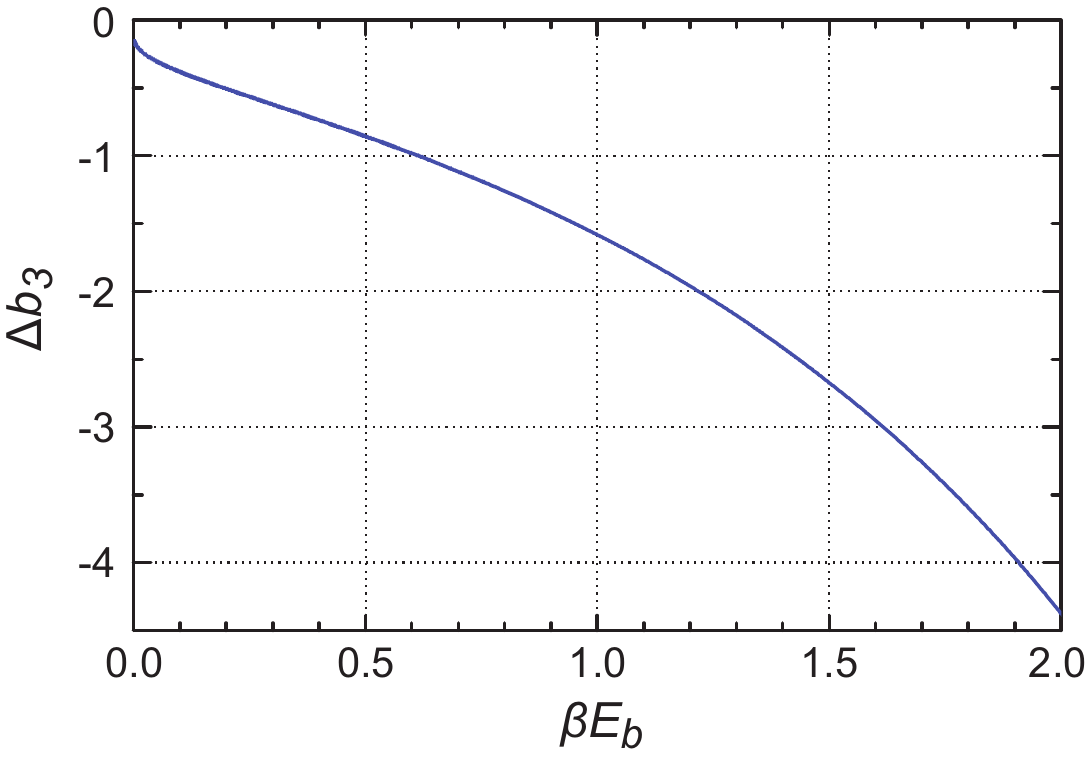} 
\end{centering}

\caption{Third order virial coefficient of a two-dimensional Fermi gas.}

\label{fig2} 
\end{figure}

The determination of the third order virial coefficient $\Delta b_{3}$ is much more involved. It was calculated in Ref. \cite{sliu10} with the help of an isotropic harmonic confinement and in Ref. \cite{sparish13} by generalizing the diagrammatic approach introduced by Leyronas \cite{sleyronas11}. In Fig.~S2, we show the numerical result. Empirically, in the range $0.003<\beta E_{b}<1$, we may use the fit, 
\begin{equation}
\Delta b_{3}=-\sum_{n=0}^{7}a_{n}x^{n},
\tag{S5}
\end{equation}
where $a_{0}=0.45938$, $a_{1}=0.40400$, $a_{2}=0.31103$, $a_{3}=0.16998$, $a_{4}=0.17801$, $a_{5}=0.23461$, $a_{6}=0.13623$ and $a_{7}=0.02685$. With the ability to calculate these coeffecients we can fit equation \ref{eq:virial_fit} to the (azimuthally averaged) density profile with $T$ and $\mu_0$ as free parameters.

\begin{center}
    \begin{table}
        \begin{tabular}{| c | c | c |}\hline
            Magnetic field (G) & $C_{OD}$ (high/low) & $C_{OD}$ (virial) \\ \hline
            972 & $ 1.27 \pm 0.08$ & $1.21 \pm 0.03$ \\ \hline
            920 & $ 1.24 \pm 0.17$ & $1.20 \pm 0.04$ \\  \hline
            880 & $ 1.20 \pm 0.12$ & $1.21 \pm 0.03$ \\ \hline
            865 & $ 1.24 \pm 0.13$ & $1.22 \pm 0.02$ \\ \hline
        \end{tabular}
        \caption{Correction factors applied to the measured optical density of the 2D clouds at each magnetic field used in the experiment. The first estimate of the correction factor is given by $C_{OD}$ (high/low) which is determined by the ratio of the optical densities in both high intensity and low intensity images. The correction factor listed is the average value of $C_{OD}$ in the range and the uncertainty is given by the standard error. The values listed in column $C_{OD}$ (virial) give the scaling factor which minimises the difference between the measured density and density determined by fits to the virial expansion. Error estimates for these measurements is determined by the 95\% confidence interval of the fit.}
    \label{tab:supp_od_correction}
    \end{table}
\end{center}

\begin{figure}[h!]
    \centering
    \includegraphics[clip,width=0.45\textwidth]{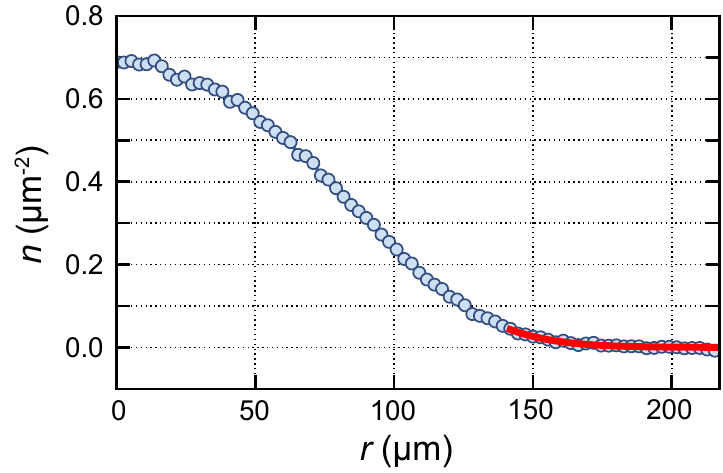}
    \caption{Fit of the in-trap density with the third-order virial expansion. Blue points are the average of 20 experimental images. Red solid line is the virial expansion fit. After convergence of the bisection algorithm the value of $\beta E_b = 0.005(1)$ is found at an optimised $C_{OD}$ of 1.21(2).}
    \label{fig:supp_2_virial_fit}
\end{figure}

The fitting procedure is sensitive to the absolute temperature $T$ via the $1/\lambda^2$ term and also through $\beta E_b$ dependence of the virial coefficients. Additionally, any errors in the correction factor $C_{OD}$ can lead to systematic errors in the fit. To account for each of these effects, we perform the fitting procedure iteratively using a bisection algorithm for different values of $\beta E_b$ until the value of $\beta E_b$ used to determine the virial coefficients converges with the fitted temperature $T$. We also perform this fitting procedure with different values of the optical density correction factor $C_{OD}$ to determine the combination of $\beta E_b, \, T$ and $C_{OD}$ which gives the best fit to the measured density. This allows us to further refine the value of $C_{OD}$ as the quality of the fit is quite sensitive to the overall scaling factor. The right hand column of table \ref{tab:supp_od_correction} shows the optimised values of $C_{OD}$ that give the best fits at each magnetic field. The uncertainty in $C_{OD}$ is reduced to the few percent level by this procedure. In Fig.~\ref{fig:supp_2_virial_fit}, we show an example of the resulting fit of the virial expansion to the cloud wings for $\beta E_b = 0.005(1)$.

At this point we have a first estimate of the absolute temperature and chemical potential of the 2D cloud which provides a basis upon which to build up the full thermodynamic description.

\subsection{Full thermodynamic analysis and validation of virial fits}

Equations (4) and (5) in the main paper provide a general approach to evaluate the relative temperature $\tilde{T} = T/T_F$ and chemical potential $\beta \mu$ via integration of the $\tilde{\kappa}$ vs.~$\tilde{p}$ equation of state at any temperature or interaction strength. Having performed the virial fit as described above, we can use the virial expansion to calculate $\tilde{\kappa}$ and $\tilde{p}$ at the $\beta E_b$ determined from the fit using
\begin{equation}
   \kappa  = \frac{2\beta}{\left(n\lambda\right)^{2}}\left[\frac{\textit{\yogh}}{1+\textit{\yogh}}+4\Delta b_{2}\textit{\yogh}^{2}+9\Delta b_{3}\textit{\yogh}^{3}+\cdots\right],
\tag{S6}
\label{eq:virial_k}
\end{equation}
\begin{equation}
   P = \frac{2}{\beta\lambda^{2}}\left[\int\limits _{0}^{\infty}dt\ln\left(1+\textit{\yogh}e^{-t}\right)+\Delta b_{2}\textit{\yogh}^{2}+\Delta b_{3}\textit{\yogh}^{3}+\cdots\right].
\tag{S7}
\label{eq:virial_p}
\end{equation}
While these are valid only in the low density wings of the cloud where the relative temperature is high, they provide useful initial conditions for evaluating the definite integrals for $\tilde{T}$ and $\beta \mu$, equations (4) and (5). Rearranging the expression for the density, equation (S2), we find that $\tilde{T}$ is given by 
\begin{equation}
\tilde{T}=\left[\ln\left(1+\textit{\yogh}\right)+2\Delta b_{2}\textit{\yogh}^{2}+3\Delta b_{3}\textit{\yogh}^{3}+\cdots\right]^{-1}.
\tag{S8}
\end{equation}

\begin{figure*}[t!]
    \centering
    \includegraphics[clip,width=0.7\textwidth]{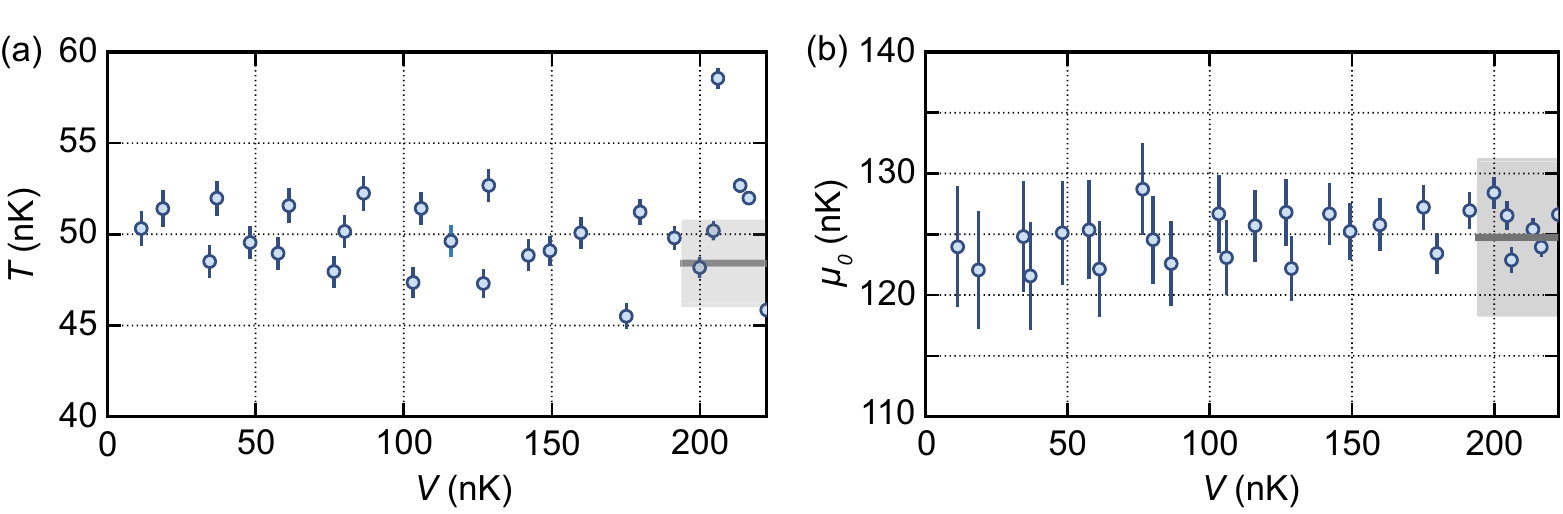}
    \caption{Result of the temperature and chemical potential validation checks for a cloud with $\beta E_b = 0.005(1)$. Panel (a) shows the resulting bulk temperature $T$ calculated from different points in the cloud, whereas panel (b) is the calculated central chemical potential $\mu_0$. Solid grey lines represents the value of $T$ and $\mu_0$ found via the fit to the wings of the in-trap density using the virial expansion and the shaded grey bands are the 95\% confidence intervals of the fit.}
    \label{fig:supp_temp_mu_check}
\end{figure*}

Under the assumptions of thermal equilibrium and the Local Density Approximation (LDA) we require that the absolute temperature across the cloud be uniform and that the chemical potential evaluated from equation (5) should satisfy $\mu(\mathbf{r}) = \mu_0 - V(\mathbf{r})$. To test compliance with these criteria, we first use the temperature integral, equation (4), to find the relative temperature $\tilde{T}$ for each value of $\tilde{p}$ in the $\tilde{\kappa}$-$\tilde{p}$ equation of state. From this we can find the absolute temperature $T$ using the local Fermi temperature $T_F(\mathbf{r}) = E_F(\mathbf{r})/k_B$, with $E_F(\mathbf{r}) = (\pi \hbar^2 / m) n(\mathbf{r})$ being set by the density. Figure \ref{fig:supp_temp_mu_check}(a) shows the absolute temperature $T$ determined at various positions through the cloud by varying the endpoint of the integration equation (4). The data are approximately uniform across the entire cloud with some fluctuations appearing due to noise in the density profile $n(\mathbf{r})$. The absolute temperature is consistent with the virial result, grey line, where the uncertainty in the virial fit is indicated by the light grey band. Averaging the values in Fig.~\ref{fig:supp_temp_mu_check}(a) over the different densities provides a robust estimate of the absolute temperature with a smaller uncertainty than that resulting from the virial fit. These provide the final values of $\beta E_b$ which are given for each magnetic field in table \ref{tab:supp_beta_eb}.

A similar check can be performed to ensure the consistency of the chemical potential. It follows from the LDA that one can determine $\mu_0$ from the value of $\mu(\mathbf{r})$ found at any position through the cloud using equation (5) from the main text. Again the virial expansion provides the initial values $(\beta \mu)_i$ and $\tilde{T}_i$ from which $\beta \mu$ at any point in the cloud can be determined. As $\beta \mu$ is calculated for each $\tilde{p}$ we can infer the chemical potential in the trap centre using the temperature determined above and our knowledge of the trapping potential. Plotting the $\mu_0$ determined in this way allows us to validate both the LDA and the thermometry. Figure \ref{fig:supp_temp_mu_check}(b) shows a plot of $\mu_0$ found by terminating the integral in equation (5) at different positions through the cloud. The overall flatness of the curve confirms the consistency of the parameters found and used in the analysis. The fact that our results are fully consistent with the virial expansion in the high temperature limit provides a rigorous confirmation of the validity of the approach.

\begin{center}
    \begin{table}
    \begin{tabular}{| c | c | c |}\hline
        Magnetic field (G) & $\beta E_b$ (virial) & $\beta E_b$ (full analysis)  \\ \hline
        972 & $0.005(1)$ & $0.005(1)$ \\ \hline
        920 & $0.06(1)$ & $0.06(1)$ \\  \hline
        880 & $0.29(5)$ & $0.26(2)$ \\ \hline
        865 & $ 0.49(5)$ & $0.47(3)$ \\ \hline
    \end{tabular}
    \caption{Values of $\beta E_b$ determined at each magnetic field. $\beta E_b$ (virial) is found by via the fitting to the in-trap density distribution with the third order virial expansion after finding the optimal $C_{OD}$. The quoted uncertainties represent the 95\% confidence interval of the fit. We also show the value of $\beta E_b$ as determined via the full thermodynamic analysis using the dimensionless compressibility and pressure. In this case the error includes both statistical uncertainty and systematic uncertainty arising due to measured field strengths and trapping frequency.}
    \label{tab:supp_beta_eb}
    \end{table}
\end{center}

\subsection{Additional thermodynamic parameters} 

\begin{figure*}[t!]
    \centering
        \includegraphics[clip,width=0.7\textwidth]{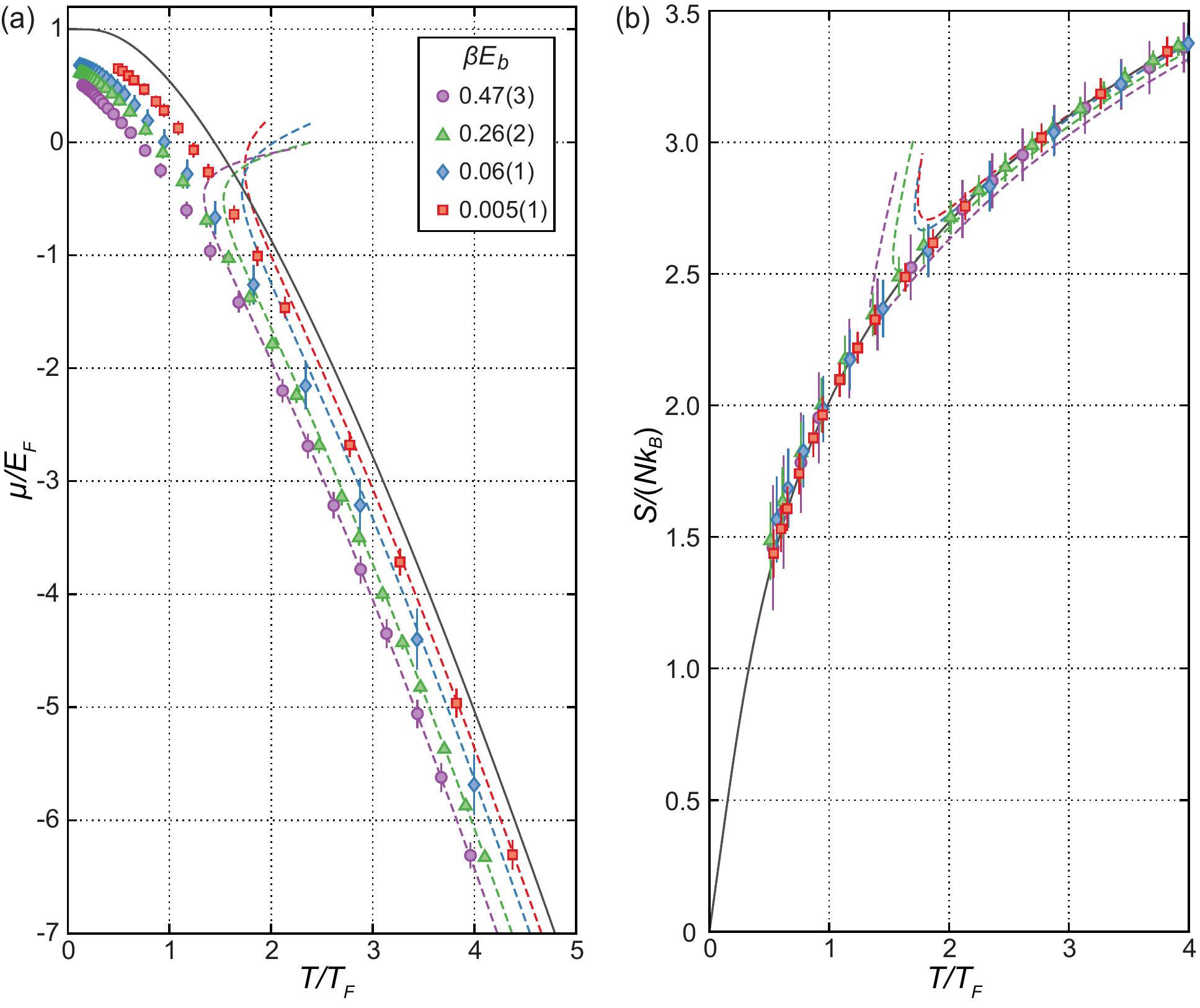}
        \caption{(a) Chemical potential and (b) entropy per particle of a two-dimensional Fermi gas with attractive interactions. The chemical potential can be found from images of a single cloud but the entropy requires differentiation of the free energy across clouds with different $a_{2D}$. In both cases the solid lines show the ideal gas result and the dashed lines are the virial expansion for the corresponding interaction strengths.}  
        \label{fig:supp_5}
\end{figure*}

Having determined $\beta \mu$ and $T/T_F$, we can take the product of these to find the dimensionless chemical potential $\mu / E_F$ which, in the case of a homogenous gas, is equivalent to the dimensionless Gibbs free energy. This is plotted in Fig.~\ref{fig:supp_5} for the four different interaction strengths. As can be seen, the chemical potential for interacting gases remains monotonic in the range of the normal phase that we measure but lies well below the ideal gas result even for modest interactions.

Unlike the 3D Fermi gas at unitarity \cite{sku12}, where the pressure follows a simple relation $P = 2/3 \, \mathcal{E}$ where $\mathcal{E}$ is the energy density, it is not possible to determine parameters such as the internal energy $U$ or entropy $S$ from measurements on a single (interacting) 2D cloud. Moreover, other variables such as Tan's contact parameter $\mathcal{C}$ require derivatives with respect to the interaction parameter, which can only be obtained through combining measurements of clouds prepared with different interaction strengths $\ln{(k_F a_{2D})}$ where $k_F = \sqrt{2 \pi n}$ is the Fermi wavevector and $a_{2D}$ is the 2D scattering length. As described in the main text, the Helmholtz free energy $F$ is given by $\tilde{F} = F/(NE_F) = \tilde{p} - \frac{1}{2}\mu/E_F$. Using the Tan relations for the contact density in 2D $\mathcal{C} = d\tilde{F} / (d \ln a_{2D})$  \cite{shofmann12,swerner12,svaliente12} one can find the average energy per particle $U/(NE_F) = \tilde{p}/2 - \mathcal{C}/k_F^4$. With this it is possible to determine further thermodynamic parameters provided the experiment is repeated at multiple values of $a_{2D}$. As we are required to differentiate with respect to $a_{2D}$ at fixed $\tilde{T}$ and we only have data at four unique values of $a_{2D}$ we necessarily introduce additional uncertainties in the derived quantities when compared to parameters such as chemical potential and temperature which can be obtained from a single cloud. This is especially important for the extreme data sets and such measurements could be improved with additional data covering a larger range of $a_{2D}$. However, as the overall contribution of the contact is small for the range of interaction strengths covered in these experiments we still find relatively small error bars and good agreement with the virial expansion at high temperatures.

Having obtained $U/(NE_F)$ it is straightforward to extract the entropy per particle $S/(Nk_B)$ from the Helmholtz free energy. In Fig.~\ref{fig:supp_5}(b) we plot the entropy in this way. It is interesting that the measured entropy differs only very slightly from the ideal gas entropy for all interaction strengths we consider. Deviations from the ideal gas entropy may become more significant at lower temperature, particularly below the superfluid transition.


\begin{thebibliography}{10}
\bibitem{mermin66} N. D. Mermin, and H. Wagner, Phys. Rev. Lett. {\bf 17}, 1133 (1966).  
\bibitem{hohenberg67} P. C. Hohenberg, Phys. Rev. {\bf 158}, 383 (1966).  
\bibitem{dagotto94} E. Dagotto, Rev. Mod. Phys. {\bf 66}, 763 (1994).
\bibitem{geim07} A. K. Geim, and K. S. Novoselov, Nature Mat. {\bf 6}, 183 (2007).
\bibitem{hasan10} M. Z. Hasan, and C. L. Kane Rev. Mod. Phys. {\bf 82}, 3045 (2010).
\bibitem{bloch08} I. Bloch, J. Dalibard, and W. Zwerger, Rev. Mod. Phys. {\bf 80}, 885 (2008).
\bibitem{bcsbecbook12} W. Zwerger, (ed.), BCS-BEC Crossover and the Unitary Fermi Gas (Lecture Notes in Physics, Springer, 2012).
\bibitem{levinsen15} J. Levinsen, and M. M. Parish, Annual reviews of cold atoms and molecules {\bf 3}, 1 (2015).  
\bibitem{modugno03} G. Modugno, F. Ferlaino, R. Heidemann, G. Roati, and M. Inguscio, Phys. Rev. A {\bf 68}, 011601 (2003).  
\bibitem{martiyanov10} K. Martiyanov, V. Makhalov, and A. Turlapov, Phys. Rev. Lett. {\bf 105}, 030404 (2010).  
\bibitem{dyke11} P. Dyke, E. D. Kuhnle, S. Whitlock, H. Hu, M. Mark, S. Hoinka, M. Lingham, P. Hannaford, and C. J. Vale, Phys. Rev. Lett. {\bf 106}, 105304 (2011).  
\bibitem{frohlich11} B. Fr\"{o}hlich, M. Feld, E. Vogt, M. Koschorreck, W. Zwerger, and M. K\"{o}hl, Phys. Rev. Lett. {\bf 106}, 105301 (2011).  
\bibitem{sommer12} A. T. Sommer, L. W. Cheuk, M. J. H. Ku, W. S. Bakr, and M. W. Zwierlein, Phys. Rev. Lett. {\bf 108}, 045302 (2012).  
\bibitem{feld11} M. Feld, B. Fr\"{o}hlich, E. Vogt, M. Koschorreck, and M. K\"{o}hl, Nature {\bf 480}, 75 (2011).
\bibitem{koschorreck12} M. Koschorreck, D. Pertot, E. Vogt, B. Fr\"{o}hlich, M. Feld, and M. K\"{o}hl, Nature {\bf 485}, 619 (2012).
\bibitem{zhang12} Y. Zhang, W. Ong, I. Arakelyan, and J. E. Thomas, Phys. Rev. Lett. {\bf 108}, 235302 (2012).  
\bibitem{ries14} M. G. Ries, A. N. Wenz, G. Z\"{u}rn, L. Bayha, I. Boettcher, D. Kedar, P. A. Murthy, M. Neidig, T. Lompe, and S. Jochim, Phys. Rev. Lett. {\bf 114}, 230401 (2015).  
\bibitem{murthy15} P. A. Murthy, I. Boettcher, L. Bayha, M. Holzmann, D. Kedar, M. Neidig, M. G. Ries, A. N. Wenz, G. Z\"{u}rn, and S. Jochim, Phys. Rev. Lett. {\bf 115}, 010401 (2015). 
\bibitem{makhalov14} V. Makhalov, K. Martiyanov, and A. Turlapov, Phys. Rev. Lett. {\bf 112}, 045301 (2014).  
\bibitem{petrov03} D. S. Petrov, M. A. Baranov, and G. V. Shlyapnikov, Phys. Rev. A {\bf 67}, 031601(R) (2003).  
\bibitem{martikainen05} J.-P. Martikainen, and P. T\"{o}rm\"{a}, Phys. Rev. Lett. {\bf 95}, 170407 (2005).
\bibitem{botelho06} S. S. Botelho, and C. A. R. S\'{a} de Melo, Phys. Rev. Lett. {\bf 96}, 040404 (2006).
\bibitem{zhang08} W. Zhang, G. D. Lin, and L.-M. Duan, Phys. Rev. A {\bf 78}, 043617 (2008).  
\bibitem{fischer14} A. M. Fischer, and M. M. Parish, Phys. Rev. B {\bf 90}, 214503 (2014).
\bibitem{bertaina11} G. Bertaina, and S. Giorgini, Phys. Rev. Lett. {\bf 106}, 110403 (2011).
\bibitem{bauer14} M. Bauer, M. M. Parish, and T. Enss, Phys. Rev. Lett. {\bf 112}, 135302 (2014).  
\bibitem{anderson15} E. R. Anderson, and J. E. Drut, Phys. Rev. Lett. {\bf 115}, 115301 (2015).
\bibitem{petrov01} D. S. Petrov, and G. V. Shlyapnikov, Phys. Rev. A {\bf 64}, 012706 (2001).
\bibitem{idziaszek05} Z. Idziaszek, and T. Calarco, Phys. Rev. A {\bf 71}, 050701(R) (2005).  
\bibitem{kestner06} J. P. Kestner and L.-M. Duan, Phys. Rev. A {\bf 74}, 053606 (2006).
\bibitem{haller10} E. Haller, M. J. Mark, R. Hart, J. G. Danzl, L. Reichs\"{o}llner, V. Melezhik, P. Schmelcher, and H.-C. N\"{a}gerl, Phys. Rev. Lett. {\bf 104}, 153203 (2010).
\bibitem{sala12} S. Sala, P.-I. Schneider, and A. Saenz, Phys. Rev. Lett. {\bf 109} 073201 (2012).
\bibitem{dyke14} P. Dyke, K. Fenech, T. Peppler, M. G. Lingham, S. Hoinka, W. Zhang, B. Mulkerin, H. Hu, X.-J. Liu, and C. J. Vale, Phys. Rev. A {\bf 93}, 011603(R) (2016). 
\bibitem{fischer13} A. M. Fischer, and M. M. Parish, Phys. Rev. A {\bf 88}, 023612 (2013).  
\bibitem{ku12} M. J. H. Ku, A. T. Sommer, L. W. Cheuk, and M. W. Zwierlein, Science {\bf 335}, 563 (2012). 
\bibitem{desbuquois14} R. Desbuquois, T. Yefsah, L. Chomaz, C. Weitenberg, L. Corman, S. Nascimb\`{e}ne, and J. Dalibard, Phys. Rev. Lett. {\bf 113}, 020404 (2014).
\bibitem{ho04} T.-L. Ho, Phys. Rev. Lett. {\bf 92}, 090402 (2004).
\bibitem{vanhouke12} K. Van Houcke, F. Werner, E. Kozik, N. Prokof'ev, B. Svistunov, M. J. H. Ku, A. T. Sommer, L. W. Cheuk, A. Schirotzek, and M. W. Zwierlein, Nature Phys. {\bf 8}, 366 (2012).
\bibitem{dalfovo99} F. Dalfovo, S. Giorgini, L. P. Pitaevskii, and S. Stringari, Rev. Mod. Phys. {\bf 71}, 463 (1999). 
\bibitem{smith05} N. L. Smith, W. H. Heathcote, G. Hechenblaikner, E. Nugent and C. J. Foot, J. Phys. B {\bf 38}, 223 (2005).
\bibitem{rath10} S. P. Rath, T. Yefsah, K. J. G\"{u}nter, M. Cheneau, R. Desbuquois, M. Holzmann, W. Krauth, and J. Dalibard, Phys. Rev. A {\bf 82}, 013609 (2010).
\bibitem{supplement} See Supplemental Material, which includes Refs.~\cite{leyronas11,ngampruetikorn13}, for further information on cloud preparation, calibration of absorption imaging and full details of data analysis.
\bibitem{leyronas11} X. Leyronas, Phys. Rev. A {\bf 84}, 053633 (2011).
\bibitem{ngampruetikorn13} V. Ngampruetikorn, J. Levinsen, and M. M. Parish, Phys. Rev. Lett. {\bf 111}, 265301 (2013).
\bibitem{reinaudi07} G. Reinaudi, T. Lahaye, Z. Wang, and D. Gu\'{e}ry-Odelin, Opt. Lett. {\bf 32}, 3143 (2007).
\bibitem{yefsah11} T. Yefsah, R. Desbuquois, L. Chomaz, K. J. G\"{u}nter, and J. Dalibard, Phys. Rev. Lett. {\bf 107}, 130401 (2011).
\bibitem{chomaz12} L. Chomaz, L. Corman, T. Yefsah, R. Desbuquois, and J. Dalibard, New J. Phys. {\bf 14}, 055001 (2012).
\bibitem{hung13} C.-L. Hung, and C. Chin, in Quantum gas experiments: Exploring many-body states, P. T\"{o}rm\"{a}, and K. Sengstock (ed.), (World Scientific 2014).
\bibitem{liu10} X.-J. Liu, H. Hu, and P. D. Drummond, Phys. Rev. B {\bf 82}, 054524 (2010).
\bibitem{nascimbene10} S. Nascimb\`{e}ne, N. Navon, K. J. Jiang, F. Chevy, and C. Salomon, Nature {\bf 463}, 1057 (2010).
\bibitem{hofmann12} J. Hofmann, Phys. Rev. Lett. {\bf 108}, 185303 (2012).  
\bibitem{werner12} F. Werner, and Y. Castin, Phys. Rev. A {\bf 86}, 013626 (2012).  
\bibitem{valiente12} M. Valiente, N. T. Zinner, and K. M{\o}lmer, Phys. Rev. A {\bf 86}, 043616 (2012).  
\bibitem{boettcher15} I. Boettcher, L. Bayha, D. Kedar, P. A. Murthy, M. Neidig, M. G. Ries, A. N. Wenz, G. Z\"{u}rn, S. Jochim, and T. Enss, arXiv:1509.03610 [cond-mat.quant-gas] (2015). 



\end{thebibliography}

\begin{thebibliography}{10}
\bibitem{sdyke14} P. Dyke, K. Fenech, T. Peppler, M. G. Lingham, S. Hoinka, W. Zhang, B. Mulkerin, H. Hu, X.-J. Liu, and C. J. Vale, Phys. Rev. A {\bf 93}, 011603(R) (2016). 
\bibitem{sreinaudi07} G. Reinaudi, T. Lahaye, Z. Wang, and D. Gu\'{e}ry-Odelin, Opt. Lett. {\bf 32}, 3143 (2007).
\bibitem{schomaz12} L. Chomaz, L. Corman, T. Yefsah, R. Desbuquois, and J. Dalibard, New J. Phys. {\bf 14}, 055001 (2012).
\bibitem{syefsah11} T. Yefsah, R. Desbouqouis, L. Chomaz, K. J. Gunter, and J. Dalibard, Phys. Rev. Lett {\bf 107}, 130401 (2011).
\bibitem{sliu10} X.-J. Liu, H. Hu, and P. D. Drummond, Phys. Rev. B {\bf 82}, 054524 (2010).
\bibitem{spetrov01} D. S. Petrov, and G. V. Shlyapnikov, Phys. Rev. A {\bf 64}, 012706 (2001).
\bibitem{sbloch08} I. Bloch, J. Dalibard, and W. Zwerger, Rev. Mod. Phys. {\bf 80}, 885 (2008).  
\bibitem{sparish13} V. Ngampruetikorn, J. Levinsen, J., and M. M. Parish, Phys. Rev. Lett. \textbf{111}, 265301 (2013).
\bibitem{sleyronas11} X. Leyronas, Phys. Rev. A \textbf{84}, 053633 (2011).
\bibitem{sku12} M. J. H. Ku, A. T. Sommer, L. W. Cheuk, and M. W. Zwierlein, Science {\bf 335}, 563 (2012). 
\bibitem{shofmann12} J. Hofmann, Phys. Rev. Lett. {\bf 108}, 185303 (2012).  
\bibitem{swerner12} F. Werner, and Y. Castin, Phys. Rev. A {\bf 86}, 013626 (2012).  
\bibitem{svaliente12} M. Valiente, N. T. Zinner, and K. M{\o}lmer, Phys. Rev. A {\bf 86}, 043616 (2012).  
\end{thebibliography}
\end{document}